\documentclass[sigconf]{acmart}

\usepackage{amsmath,amsfonts}
\usepackage{xcolor}

\usepackage{tabularx}
\usepackage{booktabs}
\usepackage{multirow}
\usepackage{subcaption}
\usepackage{caption}
\usepackage{acronym}
\usepackage{csquotes}
\usepackage{tikz}
\usepackage{wrapfig}
\usepackage{cleveref}
\usepackage{float}
\usepackage{ulem}

\usepackage{pgfplots}
\usepackage{pgf}
\usepgfplotslibrary{groupplots,dateplot}
\usetikzlibrary{patterns}
\usetikzlibrary{shapes, arrows, arrows.meta, positioning,shadows,trees}
\usepackage{etoolbox}

\pgfplotsset{compat=newest}

\usepackage{xspace}
\newcommand*{\eg}{e.g.\@\xspace}
\newcommand*{\ie}{i.e.\@\xspace}

\newcommand{\todo}[1]{\textbf{\color{red} TODO: #1}\xspace}
\newcommand{\change}[1]{{\color{magenta}#1}\xspace}
\newcommand\redsout{\bgroup\markoverwith{\textcolor{magenta}{\rule[0.5ex]{2pt}{0.4pt}}}\ULon}

\newcommand{\DEF}{\texttt{DEF}\xspace}
\newcommand{\RISK}{\texttt{RISK}\xspace}
\newcommand{\RRT}{\texttt{RRT}\xspace}

\newcolumntype{C}{>{\raggedright\arraybackslash}X}

\acmConference[Accepted to CODASPY '24]{Accepted to the Fourteenth ACM Conference on Data and Application Security and Privacy}{June 19--21, 2024}{Porto, Portugal}

\setcopyright{none}
\renewcommand\footnotetextcopyrightpermission[1]{}
\makeatletter
\newcommand\blfootnote[1]{%
	\begingroup
	\renewcommand{\@makefntext}[1]{\noindent\makebox[1.8em][r]#1}
	\renewcommand\thefootnote{}\footnote{#1}%
	\addtocounter{footnote}{-1}%
	\endgroup
}
\makeatother

\begin{document}

\title{From Theory to Comprehension: A Comparative Study of Differential Privacy and $k$-Anonymity}

\author{Saskia Nu\~nez von Voigt}
\email{saskia.nunezvonvoigt@tu-berlin.de}
\affiliation{%
	\institution{Technische Universit\"at Berlin}
	\city{Berlin}
	\country{Germany}
	\vspace{2ex}
}

\author{Luise Mehner}
\email{mehner@campus.tu-berlin.de}
\affiliation{%
	\institution{Technische Universit\"at Berlin}
	\city{Berlin}
	\country{Germany}
	\vspace{2ex}
}

\author{Florian Tschorsch}
\email{florian.tschorsch@tu-dresden.de}
\affiliation{%
	\institution{Technische Universit\"at Dresden}
	\city{Dresden}
	\country{Germany}
	\vspace{2ex}
}


\renewcommand{\shortauthors}{Saskia Nu\~nez von Voigt, Luise Mehner, \& Florian Tschorsch}

\begin{abstract}
	The notion of $\varepsilon$-differential privacy is a widely used concept of providing quantifiable
	privacy to individuals.
	However, it is unclear how to explain the level of privacy protection provided by a differential privacy mechanism with a set
	$\varepsilon$.
	In this study, we focus on users' comprehension of the privacy protection
	provided by a differential privacy mechanism.
	To do so, we study three variants of explaining the privacy protection provided by
	differential privacy: (1) the original mathematical definition;
	(2) $\varepsilon$ translated into a specific privacy risk; and
	(3) an explanation using the randomized response technique.
	We compare users' comprehension of privacy protection employing these
	explanatory models with their comprehension of privacy protection of $k$-anonymity
	as baseline comprehensibility.
	Our findings suggest that participants' comprehension of differential privacy
	protection is enhanced by the
    privacy risk model and the randomized response-based model.
    Moreover, our results confirm our intuition that privacy protection provided by $k$-anonymity is more comprehensible.\blfootnote{Accepted to ACM CODASPY 24, 19-21 June 2024, Porto, Portugal.}
\end{abstract}


\maketitle

\section{Introduction}
Privacy-preserving techniques
have been proposed in various domains to provide data protection guarantees.
The aim of these techniques is to minimize the risk of identifying an individual
while also maximizing the utility of the data.
One simple method is to remove or generalize attributes so that
each combination of attribute values comprises at least $k$ entries,
leading to the concept of $k$-anonymity~\cite{k-ano}.
Each individual in the data set is therefore indistinguishable from $k-1$ other individuals.
However, $k$-anonymity does not provide strong mathematical privacy guarantees,
as attribute values can be revealed in some situations~\cite{MachanavajjhalaGKV06,LiLV07}.

The privacy concept of $\varepsilon$-differential privacy~\cite{diffpri},
offers stronger privacy guarantees.
It is a mathematical definition in which randomization is used to limit the
impact on the output of an individual contributing to a database.
The privacy parameter~$\varepsilon$ determines the privacy-utility tradeoff.

It is, however, difficult for a user to comprehend the level of privacy protection provided to them resulting from a particular~$\varepsilon$.
Previous works have attempted to explain differential privacy
mechanisms~\cite{cummings,xiong},
quantify privacy guarantees~\cite{leeclifton,ecostrategy,theorybased},
and communicate privacy risks \cite{franzen,bullek}.
One approach to making the privacy parameter of differential privacy more
comprehensible is to translate $\varepsilon$ into a corresponding privacy risk,
expressed as a percentage~\cite{leeclifton,MehnerVT21}.
Another approach has used the randomized response technique~\cite{RRT} to
describe privacy protection~\cite{bullek}.
This technique involves local differential privacy, which has been shown to be
more intuitive~\cite{xiong}.
However, it is unclear whether these approaches
enhance
users' comprehension of the implications of differential privacy mechanisms~\cite{cummings}.

In contrast, for $k$-anonymity, the privacy parameter~$k$ is directly linked
to individual identifiability.
We therefore argue that $k$-anonymity is easier to understand than
privacy protection provided by differential privacy.
Based on our assumption, we investigated how we can explain the level of
privacy protection of differential privacy.
Namely, in such a way that it is possibly just as comprehensible as
$k$-anonymity.

To that end, we present three \textit{explanatory models} that
explain the privacy protection provided by differential privacy.
In each explanatory model, we use a particular translation of the
privacy parameter~$\varepsilon$ into a more intuitive concept.
These translations of~$\varepsilon$ describe the level of privacy protection,
making it easier to comprehend the implications of various differential privacy mechanisms.
We build upon existing and established strategies to communicate the
privacy protection provided by differential privacy quantitatively;
(1)~the original mathematical definition~(\DEF);
(2)~$\varepsilon$ as a privacy risk~(\RISK); and
(3)~an explanation using the randomized response technique~(\RRT).
We conducted an experimental study to investigate whether these
explanatory models enhance users' comprehension of differential privacy protection.

In our experimental study, we examined users' comprehension of the
privacy protection provided by a differential privacy mechanism compared to their
comprehension of the privacy protection provided by a $k$-anonymity mechanism.
We thus anchor the comprehension of the privacy protection of
differential privacy in general and the respective comprehensibility with each
explanatory model to the comprehensibility of $k$-anonymity.
Our comparison increases the methodological validity of our
study.
Importantly, we do not compare the two mechanisms themselves nor their level of privacy protection.
Instead, we are interested in the \textit{comprehensibility of privacy protection} provided by the mechanisms.

With our results we provide evidence that the privacy protection provided by
differential privacy is best understood using \RRT as an explanatory model.
Moreover, we establish $k$-anonymity as a baseline and an easily understandable privacy mechanism.

The paper's contribution and structure can be summarized as follows:
We present three explanatory models that include translations of the
privacy parameter to help users understand privacy protection and thus the
implications provided by a differential privacy mechanism in \Cref{sec:background}.
After designing and conducting an experimental study addressing our research questions (\Cref{sec:method}),
we performed a pilot study to validate our explanations and questions
before conducting our main study.
Our improvements designed to increase the internal validity of the
questions concerning subjective and
objective comprehension for the main study are presented in \Cref{sec:prestudy}. 
In our main study, we examined the participants' subjective and objective comprehension of the privacy protection
provided by differential privacy with the explanatory models \DEF, \RISK, and \RRT
compared to users' comprehension of the privacy protection provided by a $k$-anonymity mechanism (\Cref{sec:formalstudy}).
Lastly, we discuss limitations and future work in
\Cref{sec:discussion} and we review related work in \Cref{sec:relatedwork}.
We conclude our paper in \Cref{sec:conclusion}.

\section{Explanatory Models}
\label{sec:background}
In this section, we provide three explanatory models for the implications
of the privacy parameter $\varepsilon$ of our privacy mechanism---differential privacy.
Each model involves a translation of the privacy parameter
into a more intuitive concept.
Each translation is designed to help users
understand the level of privacy protection provided with a specified privacy parameter and thus the implications of the mechanism.
In addition, we give a brief overview of the privacy parameter of $k$-anonymity.

\subsection{Privacy Protection of $k$-anonymity}
The privacy protection of $k$-anonymity~\cite{k-ano}
relies on the concept of anonymity sets.
An anonymity set is a set of elements which are indistinguishable from each other.
The individual's entries in a database are generalized or suppressed in a way
that for each entry, there are at least $k$ entries with the same values
in all columns that might be used to re-identify an individual.
In other words, individual's entries are clustered
into anonymity sets.

The privacy parameter $k$ translates to the size of the smallest
anonymity set in the database.
The higher $k$, the more indistinguishable individuals exist in each group,
resulting in a stronger privacy protection. 
For instance, with $k=4$, the chance of correctly linking an entry of a group
to an individual is $1/4=0.25$.

\subsection{Differential Privacy Definition (\DEF)}
Differential privacy~\cite{diffpri} bounds the amount of
influence a single individual's data can have on the output of a statistical
computation over a database.
A mechanism $\mathcal{M}$ is $\varepsilon$-differentially private if for any two
neighboring data sets ($D_1$ and $D_2$), differing in one individual,
and any statistical result computed over the data sets ($S\subseteq \text{Range}(\mathcal{M})$) satisfy:
\begin{align}
	P[\mathcal{M}(D_1) \in S] \leq \mathrm{e}^{\varepsilon} P[\mathcal{M}(D_2) \in S] \text{.} \label{eq:dp}
\end{align}

The maximum distance between the probabilities of the mechanism returning the
result with each database is less than a certain quantity.
This quantity is based on the privacy parameter $\varepsilon$.
A privacy parameter closer to zero reduces the maximum distance,
which means that the amount of influence any one individual's data can have on the overall output is smaller.
A smaller privacy parameter thus yields stronger privacy protection.

The privacy parameter $\varepsilon$ translates into the factor by
which the probability of returning any other result is greater than the probability
of the same result if an individual is missing from the data set.
For instance, with $\varepsilon=\ln 3$, thus, $\mathrm{e}^{\ln3} = 3$
the probability of returning any result is at most three times
the probability of the same result if one individual is missing in the data set.

\subsection{Epsilon as Privacy Risk (\RISK)} 
Lee and Clifton~\cite{leeclifton} proposed for the Laplacian differential
privacy mechanism, a way of calculating the risk of users in a data set being identified.
In this framework, after an adversary receives an output of the differential
privacy mechanism, she then
imagines every possible scenario for a distribution of all
possible values for the individuals' data that she does not already know.
These scenarios are her so-called \textit{possible worlds}.
By comparing the probability of the mechanism returning the particular result
for each possible world, the adversary decides which possible world is most
likely to be true.
The probability of the mechanism indicating the correct possible world when
returning a result hence represents the users' risk of being identified.

Mehner et al.~\cite{MehnerVT21} simplified the framework
by assuming worst-case values for some variables, so that
the risk of being identified in a data set $p$ depends only on $\varepsilon$ and $n$:
\begin{equation}
	\label{eq:pRisk}
	p = \dfrac{1}{1+\mathrm{e}^{-\varepsilon}(n-1)}\text{,}
\end{equation}
where $n$ corresponds to the number of (unknown) possible worlds imagined by the adversary.
Thus, the privacy parameter can be translated into a privacy risk in percent.

However, the number of possible worlds $n$ may be difficult to grasp.
Moreover, $n$ depends on multiple often unspecified variables,
such as the knowledge of the adversary,
the number of individuals in the database and the number of possible values for an answer.

According to Mehner et al.~\cite{MehnerVT21}, assuming the worst-case attack scenario, an adversary might have only two possible worlds.
For example, she may be uncertain about only one individual's answer and
there may be only two possible values for that answer.
Accordingly, the worst-case value for $n=2$
resulting in the global privacy risk:
\begin{equation}
	\label{eq:globalRisk}
	p = \dfrac{1}{1+\mathrm{e}^{-\varepsilon}}\text{.}
\end{equation}

We can therefore translate the privacy parameter for a given $\varepsilon$
into the privacy risk of identifying the true answers of individuals included
in the database.
In other words, if an adversary queries the answer of an individual
and there are only two possible answer values (\ie, in a worst-case attack scenario),
we can determine the probability of the mechanism indicating the true answer of the individual for a specified $\varepsilon$.
For example, assume we set $\varepsilon=\ln 3$, which
yields a privacy risk of $75$\,\%,
\ie, in the worst-case attack scenario, the true answer of a person included in the database is revealed with a probability of $75$\,\%.


\subsection{Using Randomized Response (\RRT)}
The number of possible worlds $n$ of \Cref{eq:pRisk} is similar to the number
of different answers in the randomized response technique~\cite{RRT}.
The randomized response technique is an approach designed to provide plausible
deniability to data subjects.
The idea is that some of the data subjects will give their true answer and others will give a forced answer.
The decision of whether an individual gives a true or a forced answer is made
randomly.
Consequently, each answer has a probability of being an individual's true answer.
Therefore, users' answers do not reveal the individuals' true answers with certainty.
The randomized response technique inherently holds the local differential privacy guarantee.

More precisely, with a probability of $p_{true}$, the true answer $a$ is stored in the database.
The probability of any false answer $a'\neq a$ is
$p_{false}=(1-p_{true})/(d-1)$, where $d$ is the number of possible answers.
This mechanism is one approach of the randomized response, called unary encoding,
and it satisfies local differential privacy:
\begin{align}
	P[\mathcal{M}(a) = a ] &\leq \mathrm{e}^{\varepsilon} P[\mathcal{M}(a)=a']  \label{eq:ldp} \\
	p_{true}\mathrm{e}^{-\varepsilon} &=  p_{false} \text{,}
\end{align}
resulting in
\begin{equation}
	\label{eq:ptrue}
	p_{true} = \frac{1}{1+\mathrm{e}^{-\varepsilon}(d-1)}\text{.}
\end{equation}

The probability of storing a true answer is equal to the privacy risk~\Cref{eq:pRisk},
where the number of possible worlds $n$ corresponds to the number of different answers $d$.

Hence, we can translate the privacy parameter $\varepsilon$ into the
probability with which the mechanism stores a true answer in the database.
For example, assume we set $\varepsilon=\ln 3$ and have two possible answers~ ($d=2$).
As a result, the probability of storing the true answer is $75$\,\%.
With a higher number of possible answers, \eg $d=28$, the probability of storing
the true answer decreases to $10$\,\%.
Note that the model also works for real-valued (continuous) data.
In this case, the worst case with $d=2$ should be used.
The result indicates the probability of storing true answers,
regardless of whether the data is discrete or continuous.

\begin{figure*}[tb]
	\centering
	\begin{tikzpicture}
		\tikzstyle{every node}=[font=\scriptsize,minimum height=0.3cm,align=center, node distance=0.25cm]
		\tikzstyle{arrow}   = [solid, line width = 9pt, -{Triangle[width = 18pt, length = 12pt]}, color = black!30, shorten >= 3pt, shorten <= 3pt]
		
		\draw[arrow] (0,0) -- (15.5,0);
		
		\node[rounded corners,draw,minimum width=2.3cm,minimum height=2.5cm,fill=white] at (0,0) (step1) {};
		\node[fill=black!10] at (step1.north) {Demographics};
		\node[text width=1.7cm] at (step1) {age,\\field of study,\\current level of education};

		\node[rounded corners,draw,minimum width=2.7cm,minimum height=2.5cm,right=of step1,fill=white] (step2) {};
		\node[fill=black!10] at (step2.north) {Scenario};
		\node[text width=2.4cm,yshift=0.3cm] at (step2) {statistics drug use at school; parents should not infer their son/daughter's drug use};
		\draw[dashed] ([yshift=-0.5cm]step2.west) -- node[text width=2.4cm,below] {comprehension questions} ([yshift=-0.5cm]step2.east) ;
		
		\node[rounded corners,draw,minimum width=6.8cm,minimum height=2.5cm,right=of step2,fill=white] (step3) {};
		\node[fill=black!10] at (step3.north) {Privacy Protection Explanations (within-subject)};
		
		\node[rounded corners,draw,minimum width=4.2cm,minimum height=1.2cm,anchor=east,xshift=-0.25cm] at (step3.east) (dp){};
		\node[fill=green!10, text width=3cm] at (dp.north) {differential privacy\\(between-subject)};
		
		\node[draw,fill=black!10,yshift= -0.2cm,below=of dp.north,text width=0.5cm] (risk){\RISK};
		\node[draw,fill=black!10,right=of risk,text width=0.5cm] {\RRT};
		\node[draw,fill=black!10,left=of risk,text width=0.5cm] {\DEF};
		\node[below=0.0cm of risk,text width=4cm] {objective and subjective questions};
		
		\node[rounded corners,draw,minimum width=1.85cm,minimum height=1.2cm,left=of dp]  (kanon){};
		\node[fill=green!10] at (kanon.north) {$k$-anonymity};
		\node[text width=1.75cm] at (kanon) {objective and subjective questions};
		
		\node[fill=black!10,yshift=0.35cm,text width=4cm] at (step3.south) {direct comparison};
		
		\node[rounded corners,draw,minimum width=2.2cm,minimum height=2.5cm,right=of step3,fill=white] (step4) {};
		\node[fill=black!10] at (step4.north) {Numeracy $\ldots$};
		\node[text width=1.8cm] at (step4) {numeracy\\privacy experience\\privacy concerns};
		
		\node[rounded corners,draw,minimum height=2.5cm,minimum width=0.75cm,right=of step4,fill=white] (step5) {};
		\node[rotate=90,text width=2.8cm] at (step5) {Check question};
	\end{tikzpicture}
	\caption{Overview of the study design.}
	\label{fig:flow}
\end{figure*}
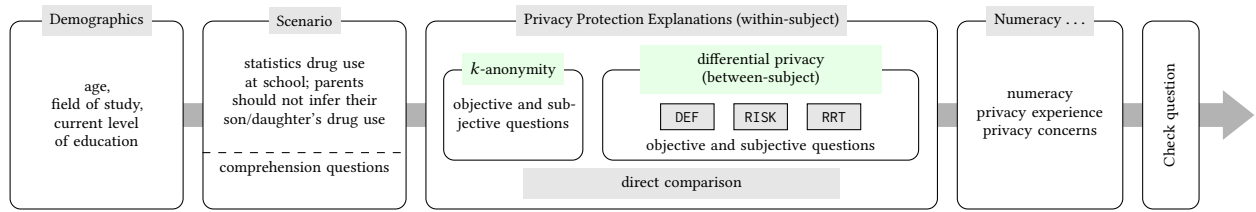

\section{Methodology} 
\label{sec:method}
In this section, we present and justify our hypotheses we formulated to design our study.
In addition, we detail how participants were instructed, describe our sample,
how we conducted the study and how we analyzed the data.


Syntactic anonymization models, such as $k$-anonymity, were originally designed for privacy-preserving data publishing~\cite{CliftonT13}.
Differential privacy, on the other hand, is more suitable for privacy-preserving data mining.
The concept of privacy-preserving data publishing usually assumes a non-expert data publisher, \ie,
the data publisher does not have the knowledge to perform data mining~\cite{FungWCY10}.
Given that $k$-anonymity is a viable solution for privacy-preserving data publishing,
the mechanism of $k$-anonymity is aimed at the non-expert who is the end user of the model.
With $k$-anonymity as as simple and intuitive model~\cite{FriedmanWS08},
we derive our first hypothesis:
\begin{itemize}
	\item[(H1)] Differential privacy vs. $k$-anonymity: The privacy protection
	provided by $k$-anonymity is easier to comprehend than the privacy protection provided by differential privacy (independent of the explanatory model).
\end{itemize}

The definition of differential privacy is complex.
Therefore, it is important to describe the techniques or the implications of a differentially private mechanism.
\RRT has often been used as an intuitive mechanism~\cite{bullek,SmartSV22}.
Previous work has~shown that \RRT provides more understanding among users~\cite{xiong,bullek}.
\RISK was developed as an intuitive explanation of $\varepsilon$.
Consequently, we derive the following hypothesis:
\begin{itemize}
	\item[(H2)] Explanatory models: The explanatory models \RRT and \RISK will provide a better comprehension of the privacy protection than the \DEF model.
\end{itemize}


Previous work has shown that both numeracy skills and level of educational affect
risk understanding~\cite{Keller09,franzen}.
Users with low numeracy skills have difficulty understanding risk in general~ \cite{Keller09}.
These findings from previous work lead us to our final hypothesis:
\begin{itemize}
	\item[(H3)] Education level and numeracy skills: High levels of education and high numeracy skills help users to comprehend the privacy protection provided by differential privacy.
\end{itemize}

\subsection{Measures}
Participants answered questions to evaluate their subjective and objective
comprehension of privacy protection.
We also included measures for covariates: demographics, privacy concerns and numeracy skills.

\subsubsection{Comprehension}
Similar to previous work~\cite{SmartSV22,xiong}, we evaluate the
subjective comprehension (perceived comprehension) and
objective comprehension (actual comprehension) of the $k$-anonymity explanation
and our explanatory models of the privacy protection provided by differential privacy~(\RISK, \RRT, and \DEF).
	
We designed the questions concerning comprehension from scratch,
using direct questions.	
We included 
three 7-point-Likert scaled questions regarding how the participants subjectively
comprehended the level of privacy protection that each mechanism (and its
respective privacy parameter) provided.
Following the questions concerning subjective comprehension, there were four questions testing the
participants' objective comprehension of privacy protection.
In addition, we gave the participants the possibility to comment on their
comprehension answers.
Last, we asked participants to directly compare which privacy mechanism they felt was most comprehensible and intuitive in terms of privacy protection.

\subsubsection{Covariates}
We assessed the participants' numeracy skills using subjective rating and objective test questions.
The numeracy questions were taken from multiple validated numeracy assessments
found in the literature~\cite{numeracyFagerlin,numeracySchrager,numeracyLipkus}.
Moreover, we asked about any previous experience with privacy mechanisms in general and
differential privacy in particular.
Finally, we also assessed the participants' general privacy concerns using
a set of questions adapted from Malhotra et al.~\cite{MalhotraKA04}.
We used these questions in the categories of collection and awareness.
We also included \enquote{attention} check questions as part of the privacy aptitude and
at the end of the study to exclude inattentive participants:
\textit{Please select 3 (More or less agree) for this question} and
\textit{What is $4+5$?}\footnote{We believe that this mathematical operation does not relate to numeracy because of its simplicity. When answered, the question was answered correctly by all participants of our main study.}.
We assume that those participants who were motivated at the end of the survey were also motivated at the beginning.
	
	
\subsection{Scenario and Explanations}
We defined a fictional scenario about drug use at school as a running example of a setting
where privacy is crucial and where privacy protection needs to be
well understood at the same time.
A school stores student answers to a questionnaire on drug use in a database grouped by age and class.
In order to raise awareness, parents can query the database, which is
protected with a privacy mechanism.

Our explanations are designed from scratch.
We used text-based explanations because we focused on evaluation of the explanatory model, not on how it was communicated.
Our explanations start with a short description of the mechanism,
inspired by the \textit{Techniques} description of Cummings
et al.~\cite{cummings}.
This was followed by an explanation of the privacy protection parameter,
\eg $k$, difference~(\DEF), risk (\RISK) and probability (\RRT).
Finally, we applied these explanations to our scenario and provided concrete examples.
The exact wording of our scenario and explanations can be found in \Cref{appendix:descriptions}.

\subsection{Experimental Process}%
Prior to the main study, we conducted a pilot study to increase the validity of our study questions.
In particular, the pilot study allowed us to validate our questions, explanations and instructions in
terms of textual clarity and general comprehensibility.
We summarize the results of the pilot study and the induced changes in \Cref{sec:prestudy}.

\subsubsection{Overview of the Study Design}
In Figure~\ref{fig:flow} we present an overview of our study design
and procedure.
The process and design of the main study and the pilot study, were the same.
Both studies had a mixed design with a between-subject factor
\enquote{explanatory model} (for differential privacy protection with three
conditions \RISK, \RRT, and \DEF)
and a within-subject factor with two levels (\enquote{privacy protection provided by $k$-anonymity} and \enquote{privacy protection of differential privacy}).
The within-subject factor included in our study allowed us to
evaluate the comprehensibility of differential privacy protection with each
explanatory model compared to
the comprehensibility of the privacy protection of $k$-anonymity.
As a results, we were able to
\begin{enumerate}
	\item verify whether the privacy protection of $k$-anonymity
	is indeed easier to comprehend than that of
	differential privacy,
	\item anchor the comprehensibility of differential privacy protection
	with each explanatory model to the comprehensibility of privacy protection of $k$-anonymity as a baseline for the best possible comprehensibility,
	\item control for any interindividual differences in comprehension skills between the three conditions.
\end{enumerate}

Moreover, use of a within-subject design reduced the standard deviation in the objective and subjective comprehension scores, improving the statistical validity of our study.

After a short welcome text explaining the purpose of the study, the
participants were asked to provide some demographic information about themselves
(age, field of study and current level of education).
Next, we introduced our fictional scenario.
We ensured that the participants understood the scenario by asking three check questions:
1) Who provides the database in the scenario? 2) What kind of data is stored in the database? 3) Eve (the adversary) wants to find out the data of whom?

\subsubsection{Procedure of Explanations}
After ensuring that the participants had read and understood the scenario,
each participant
was presented with explanations of the privacy protection of two
privacy-enhancing mechanisms,
an explanation of $k$-anonymity and an explanation of differential privacy.
To control for learning and other sequence effects, the order of the two
explanations and their respective comprehension questions were balanced.
In other words, participants were randomly assigned to either the first order group,
where the explanation and questions for $k$-anonymity were presented first,
or to the second order group, where the explanation and questions of the
differential privacy protection were presented first.
Since each participant read and answered the questions for the two explanations,
our study had a within-subject factor with the privacy protection of
differential privacy and the privacy protection of $k$-anonymity as factor levels.

The explanation of the privacy protection of $k$-anonymity was the same across all conditions.
Each participant randomly (uniformly distributed) received one of the three
explanations (\RISK, \RRT, or \DEF) for differential privacy protection,
resulting in three between-subject conditions for the factor
\enquote{explanatory model for differential privacy protection}.
We used similar phrasing and wording in all explanations, including the explanation of the privacy protection of $k$-anonymity,
in order to compare the comprehension of the explanation.
In addition, the subjective as well as the objective comprehension questions were identical for each explanation.

The level of privacy protection provided by the differential privacy mechanism,
\ie, the privacy parameter $\varepsilon$,
was the same in each explanatory model.
We wanted to rule out the possibility
of the level of privacy protection systematically interfering with the
participants' comprehension of differential privacy protection.
However, differential privacy assumes a stronger adversary than $k$-anonymity does.
A $k$-anonymity mechanism cannot provide an equally strong privacy protection as the differential privacy mechanism explained using the
\RISK, \RRT, and \DEF explanatory models in our scenario.
Therefore, we have to trust that the
weaker privacy protection did not interfere with the participants'
comprehension.
Consequently, in our study,
we explain the privacy protection of $k$-anonymity with $k=4$.
We believe that this is an appropriate value to explain the privacy protection
of $k$-anonymity since this results in a probability of being identified of $0.25$.
Again, we emphasize that we cannot match the privacy levels of the two mechanisms.

\subsubsection{Procedure after Explanations}
After providing both the explanations and the questions about their comprehensibility, we asked
participants directly about which privacy mechanism (if any) was more comprehensible with respect to
the level of privacy protection and why.
We also asked which mechanism (if any) they regarded as
providing a greater privacy protection in the particular scenario, and why.
The~latter question was implemented to gain a deeper insight into
whether the participants had gained a sense of the relationship between a
particular privacy parameter
and the respective level of privacy protection provided by each mechanism.


\subsection{Participant Recruitment and Attributes}
Both the pilot study and the main study were implemented using LimeSurvey\footnote{\url{www.limesurvey.org}}
and emailed to university students of Berlin.\footnote{We cannot exclude the possibility of participants who participated in both studies. However, the pilot study took place one year earlier, so we assume that the effect is negligible. In addition, participants were asked about their prior knowledge of privacy, so the overlap was controlled in the results of participants without any prior knowledge.}
Our main study was publicly available between February 8 and 22, 2023.
The participation was voluntary and we did not offer any remuneration.

We used a set of questions provided by the Ethics Commission of TU Berlin to
self-evaluate the ethical considerations of the planned research project.
We then decided that a detailed application to the Ethics Committee was not necessary.
However,
to address potential ethical issues, we informed the participants (of the pilot study and main study)
about our data policies in our invitation email before the survey:
The evaluation of the responses would be anonymized, \ie,
we only used the LimeSurvey Response ID
as an identifier and would remove it before the statistical analysis.
We only accessed the results of the pilot study that were necessary to validate our explanations and questions.

For the participants, the purpose of our study was to evaluate
explanations of the privacy protection provided by two privacy-enhancing mechanisms.
At that point, we did not refer to differential privacy protection as the focus of our study
to avoid the influence of demand characteristics or participant expectations
about our desired outcome of the study.
All participants were presumed to have at least a high school diploma and to be currently studying at a university.

There were a total of $249$ respondents in the main study.
Of these, only $93$ participants answered the subjective and
objective comprehension questions for both explanations and could therefore be included in the analysis.
Of these, three participants were excluded because
they gave an incorrect answer to one of the comprehension questions regarding the scenario
or because they answered one of the attention-check questions incorrectly,
resulting in a total of $90$ analyzed participants.
Of these, $78$ participants fully completed the study and thus
answered all questions.
We decided to nevertheless include the other $12$ participants who did not finish the study
into parts of our analysis to increase the statistical power of our study and to
reduce motivation bias. 
In conclusion, $90$ participants were included in analyses involving the objective and subjective comprehension,
$78$ participants were included in all our analyses, including those concerning
the direct comparison and those involving the
participants’ privacy concerns or numeracy skills.

Consequently, we included $90$ submissions in the analysis:
$27$ for \RISK,
$30$ for \RRT, and
$33$ for \DEF.
Of these participants,
$66$ indicated a \enquote{STEM} study field of science, technology, engineering, or mathematics
($28$ were students of computer science/engineering).
Five students indicated a study field of management or economics,
eight students indicated a study field related to architecture or design, and
$11$ students indicated a study field of social sciences, or psychology.
The age of the participants ranged from $18$ to $40$ years with a mean age of
approximately $25.03$ and a median age of $24$.
The level of education was high overall, with $73$ participants having a bachelor's degree or higher.
Of these, $13$ participants stated that they had a master's degree.
These $90$ participants spent an average of $34.8$ minutes on the study.

\subsection{Data Set Pre-processing and Analysis}
Each participant received a score for subjective comprehension and
an objective comprehension score, both between $0$ and $1$ corresponding
to \enquote{very poor} and \enquote{very good} comprehension, respectively.
To obtain the subjective comprehension score, we calculated the mean score of
the three subjective comprehension questions for each participant.
We thereby inverted the score of the first question so that for every question a higher score indicated greater comprehension.
We then normalized the scores to a range from $0$ to $1$.

To measure objective comprehension, we scored each correct and incorrect answer
as~$1$ and~$0$, respectively.
We calculated the mean of the four objective comprehension questions for each
participant and normalized the objective score to be between
$0$ and $1$.

In order to avoid the influence of baseline differences in the participants'
comprehension abilities between the conditions,
we calculated the differences between the comprehension scores for the
privacy protection of $k$-anonymity and differential privacy for each
participant.
In other words, we subtracted the mean scores for the objective and the subjective comprehension of the differential privacy explanation from the mean scores of the $k$-anonymity explanation model.
A positive difference means that the $k$-anonymity explanation had a higher score, and was thus easier to comprehend.
A negative difference, on the other hand, means that the differential privacy explanation has a higher score. If the difference is $0$,
the scores indicate a similar level of comprehension.

We tested for differences between the subjective comprehension scores for the two explanations for each
condition using one-tailed t-tests.
Furthermore, we examined the effect of the explanatory model for
differential privacy protection on the differences between the comprehension
scores for the two explanations.
We also examined the effect of the order of explanations, \ie, whether participants had read the explanation and answered
the corresponding comprehension questions for privacy protection of $k$-anonymity or for that of differential privacy first.
To that end, we conducted two-way
analyses of variance (ANOVAs) with two between-subject factors of \textit{explanatory model} and \textit{order}.
We conducted one ANOVA for the differences in the subjective comprehension scores and one for the differences in the objective comprehension scores.
The ANOVAs also allowed us to investigate any interaction effects between the two factors (explanatory model and order).

We also wanted to analyze how the participants’ comprehension of
differential privacy protection was influenced by privacy concerns and numeracy skills.
For that we included only participants who had completed the whole study,
as the questions regarding privacy concerns and numeracy skills were asked at the end of the study.
To test (H3), we calculated a privacy concern score as well as a subjective numeracy score and an objective numeracy score for each participant.
We used Pearson's correlation coefficient to measure the correlations.

In the final data set, each entry contains the following values:
an explanation group \{1-3\},
a mean subjective comprehension score for $k$-anonymity [0-1],
an objective comprehension score for $k$-anonymity [0-1],
a mean subjective comprehension score for differential privacy [0-1],
an objective comprehension score for differential privacy [0-1],
a comparison regarding comprehensibility \{$k$-anonymity, differential privacy, both\},
a comparison regarding prevention \{$k$-anonymity, differential privacy, both\},
education \{high school, bachelor, master\},
a privacy concerns score [1-7],
a subjective numeracy score [0-2] and
an objective numeracy score~[0-8].

\subsection{Limitations}
Our study was conducted with students from German universities only.
Therefore, our results cannot be transferred to the general public.
To allow more generic inferences, future work should test a more
heterogeneous sample.
Also, the number of participants was limited and many participants aborted the study.
As a result, the statistical power might simply not have been sufficient to detect all effects.
Therefore, future work should investigate a higher number of participants, or
should execute a power analysis, using our effect sizes as a basis.
To increase the statistical power of our study and to reduce motivation
bias, we indeed included participants who answered questions concerning subjective and
objective comprehension for both explanations, but did not fully complete the
study. However, these responses were only used in certain parts of our analysis.
Although these participants answered the scenario-check question,
they did not answer our attention- and math-check questions at the end of our study.
Therefore, we cannot be certain that those participants answered our study attentively.

Furthermore, we compared the comprehensibility of different privacy mechanisms,
which also provided different levels of privacy protection.
Therefore, we cannot rule out that this difference had an effect on our results.
However, it is unlikely as we did not compare the privacy protection provided by the mechanisms but only the participants' comprehension.

Most importantly, our objective comprehension questions may have been
inherently easier to answer for one of the mechanisms' privacy protection or
for one of the explanatory models for differential privacy protection.
Even though we conducted a pilot study to validate our scenario,
our explanations, and our comprehension questions in terms of textual clarity and general comprehensibility,
we had to apply some adjustments to the objective comprehension questions.
As a result, the modified questions were not validated before the main study.
Therefore, the objective comprehension questions may have inherently favored one of the explanations.

\section{Pilot Study}
\label{sec:prestudy}
The primary goal of the pilot study was to evaluate our study questions
with respect to ambiguity, difficulty, and internal consistency.
Furthermore, the pilot study allowed us to refine the wording of the questions, explanations, and study instructions.
Moreover, it allowed us to consider and address the comments provided by the participants.
These comments were read by one author and selected if they included feedback about the instructions\footnote{The comments were solely intended to improve the wording of the questions, the explanations, and the study instructions, and were not primary research artifacts. Therefore, we did not use any further statistical methods for these entries.}.

We had the following findings from our pilot study.
The majority of incorrect values for our check question were very close to the correct value.
Hence, some of the answers may have been incorrect due to mathematical difficulties rather than inattentiveness.
Therefore, we replaced the \enquote{attention} check question (\textit{What is $15+7$?}) with a simpler calculation
to reach less mathematically able people: \textit{What is $4+5$?}

All of our explanations were generally understandable, indicated by high mean comprehension scores for all explanations~($\text{min}=0.62$, $\text{max}=0.92$).
In the main study, we also recorded the order of explanations for $k$-anonymity and
differential privacy,
to derive any possible relationship between the comprehension and the order of the explanations.

The means of the subjective scores were similar across all conditions.
Therefore,
we only aligned the wording so that all questions explicitly enquired about the \textit{level of privacy protection} instead of about the mechanism itself.

We removed one objective comprehension question
since this question was only answered correctly by a very few participants
compared to the other questions.
In addition, we modified objective questions~\#1--\#3
to maintain consistency in wording, by adding the concept of the \textit{privacy parameter} in the explanations and in question~\#2.
In question~\#3, we asked for the implications when the \textit{privacy parameter is $0$}.

The comments overall suggested a high level of comprehension of privacy protection for both mechanisms.
We also confirmed that the privacy protection of $k$-anonymity was indeed regarded to be as inherently easier to comprehend, and a less complex privacy mechanism.
The comments led us to shorten the explanations as much as possible and to explain that the aim of the study focused on the comprehension of privacy protection rather than of the mechanisms themselves.
The comments also led us to focus more on the link between the privacy parameter and the
level of privacy protection provided, and less on how the mechanisms work.

Many participants were confused about the phrase \enquote{random noise} in the
explanation used in our pilot study, and the specified results of the differential privacy mechanism.
We decided to avoid mathematical vocabulary wherever possible,
to completely exclude the concept of random noise from the explanations and to omit any specific calculations or results returned by the mechanism.
Instead, we emphasize the probabilities and describe that
differential privacy provides privacy protection
by randomly modifying statistical results.
We also received indications that it would be helpful to include the information in our scenario that Eve, the adversary,
knows that the returned result may not be correct.
Furthermore, we changed the wording of the sample database answers from \enquote{true/false} to \enquote{yes/no}, because this seemed to be confusing
in light of the \enquote{true answers of the students}.

\begin{table}[tb]
	\caption{Mean comprehension scores for the three explanatory models}
	\label{tab:mean_tables}
	\setlength{\tabcolsep}{4pt}
	\begin{tabularx}{\columnwidth}{p{0.7cm}p{0.7cm}XXXX}
		\toprule
		& & \multicolumn{2}{c}{$k$-anonymity} & \multicolumn{2}{c}{differential privacy} \\
		\cmidrule(lr){3-4}\cmidrule(lr){5-6}
		& N & subjective & objective & subjective & objective \\
		\midrule
		\RISK & 27 & 0.78 & 0.77 & 0.64 & 0.51 \\
		\RRT &  30 & 0.70 & 0.73 & 0.65 & 0.50 \\
		\DEF &  33 & 0.78 & 0.76 & 0.45 & 0.43 \\
		\bottomrule
	\end{tabularx}
\end{table}

\section{Results}
\label{sec:formalstudy}
In the following, we describe the results of our main study.
Throughout our analysis, we use a significance-level of $\alpha=$~$0.05$ and adjust the results of the post-hoc t-tests with
Bonferroni corrections.\footnote{The false positive error grows with the number of tests performed. A common approach to deal with this is the Bonferroni correction, which sets
	$\alpha$ for the entire set of $n$ comparisons equal to $\alpha=\alpha/n$. For example, if we have a set of three hypothesis tests and $\alpha=0.05$, our adjusted significance level equals $0.05 / 3 = 0.017$.}


\subsection{Subjective Comprehension}
\label{sec:subjective}
The mean scores of subjective comprehension in \Cref{tab:mean_tables} show~that
across all explanatory models for differential privacy protection,
the level of privacy protection resulting from $k$-anonymity was easier to
comprehend than that resulting from differential privacy.
One-tailed t-tests revealed a significant difference in comprehensibility,
with higher scores for the privacy protection of $k$-anonymity,
using the \RISK model~($t \approx 4.522$, $p < 0.001$, Cohens $d \approx 0.762$)
and using the \DEF model ($t \approx 7.586$, $p <0.001$, Cohens $d \approx 1.749$).
These results support (H1).
In contrast to (H1), when we used the \RRT model,
the difference in comprehension between the privacy protection of $k$-anonymity and differential privacy was not significant for subjective comprehension
($t \approx 1.27$,~$p \approx 0.107$).

\begin{figure}[tb]
	\begin{subfigure}[t]{.45\columnwidth}
		\centering
		\includegraphics[width=\columnwidth]{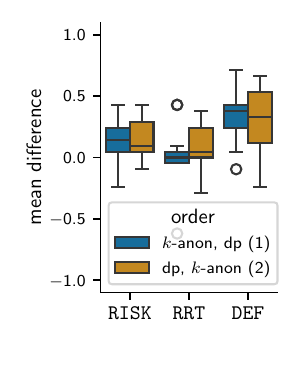}
		\vspace{-2em}
		\caption{Subjective}
		\label{fig:boxplot-subjective}
	\end{subfigure}%
	\begin{subfigure}[t]{.45\columnwidth}
		\centering
		\includegraphics[width=\columnwidth]{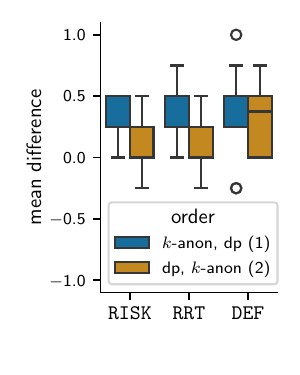}
		\vspace{-2em}
		\caption{Objective}
		\label{fig:boxplot-objective}
	\end{subfigure}
	\caption{Differences between scores for comprehension of $k$-anonymity and differential privacy.}
	\label{fig:boxplotDifference}
	\vspace{-1em}
\end{figure}

The subjective comprehension scores regarding differential privacy protection were
higher when using the \RISK model than the \DEF model, and slightly higher when using the \RRT model.
We present the difference between the mean subjective scores of $k$-anonymity
and the mean subjective score of differential privacy
in \Cref{fig:boxplot-subjective}.
In \RRT the differences were distributed around zero, with a median
of zero, whereas the interquartile range and median in \RISK were
above zero, but lower than in \DEF.

We tested the distribution of the differences with \RISK, \RRT,~and \DEF for each order of explanations with D'Agostino's K-squared tests.
There was no indication of the differences not being normally distributed for any of the conditions.
Furthermore, Levene's test did not show any significant differences of the variances between any of the conditions,
indicating equality of variances.
Hence, all requirements for conducting a between-subject two-way ANOVA were met.
The ANOVA revealed a significant effect of the explanatory model on the difference between the subjective comprehension scores for the privacy protection of $k$-anonymity and differential privacy, with $F \approx 12.979$,
$p<0.001$ and $\eta^{2} \approx 0.234$.
There was no significant effect of the order of explanations on the differences.
Furthermore, there was no significant interaction between explanatory model and order of explanations.

One-tailed t-tests showed that in comparison to the \DEF model,
the mean difference was significantly smaller with the \RISK model ($t \approx -3.342$, $p<0.001$, Cohens~$d \approx0.809$)
as well as with the \RRT model ($t \approx-4.624$, $p<0.001$, Cohens $d\approx1.166$).
Moreover, the difference in comprehensibility was smaller when we used the \RRT model than when we used the \RISK model ($t\approx-1.725$, $p\approx0.045$, Cohens $d\approx0.458$).
These results corroborate (H2), where we hypothesized that the
comprehension of the \RISK and the \RRT model is higher than of the \DEF model.
For the \RRT model, participants
achieved subjective comprehension scores comparable to those for the
privacy protection of $k$-anonymity, a privacy mechanism that
is supposedly less complex and whose privacy protection is more intuitively understandable.

\subsection{Objective Comprehension}
\label{sec:objective}

From \Cref{tab:mean_tables}, we infer that the objective comprehension scores 
for $k$-anonymity were higher than for the differential privacy explanation.
The scores for the objective comprehension of the privacy protection of
differential privacy were generally low for all explanations.
With a mean of around $0.5$, the scores correspond to the expected success rate by randomly guessing the answers.

\begin{figure}[tb]
	\begin{subfigure}[t]{.45\columnwidth}
		\centering
		\includegraphics[width=\columnwidth]{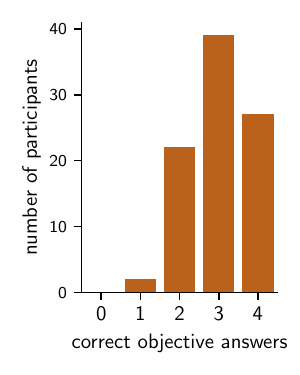}
		\caption{$K$-anonymity}
		\label{fig:objectiveK}
	\end{subfigure}%
	\begin{subfigure}[t]{.45\columnwidth}
		\centering
		\includegraphics[width=\columnwidth]{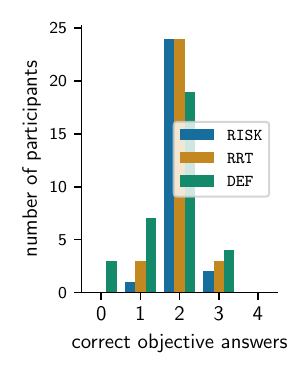}
		\caption{Differential privacy}
		\label{fig:objectiveDP}
	\end{subfigure}
	\caption{Proportion of correctly answered questions on objective comprehension.}
	\label{fig:objective}
	\vspace{-1em}
\end{figure}

In \Cref{fig:objective}, we show the number of correct objective answers for
each explanatory model.
For $k$-anonymity, all participants had at least one correct answer (cf. \Cref{fig:objectiveK}).
As illustrated in \Cref{fig:objectiveDP}, three participants answered all objective comprehension questions wrong.
All three participants were part of the \DEF group.
Notably, none of the participants answered all the questions correctly.
One-tailed t-tests revealed a significant difference in comprehensibility,
with higher scores for the privacy protection of $k$-anonymity,
with \RISK ($t \approx 6.31$, $p < 0.001$, Cohens~$d \approx 1.72$),
with \RRT ($t \approx 6.18$, $p <0.001$, Cohens~$d \approx 1.59$)
and with \DEF ($t \approx 6.62$, $p <0.001$, Cohens~$d \approx 1.51$).
These results confirm~(H1).

In \Cref{fig:boxplot-objective}, we plot the mean differences of scores between
$k$-anonymity and differential privacy.
The mean difference between the objective comprehension of privacy protection
provided by $k$-anonymity and differential privacy
was smallest with the \RRT model, followed by the \RISK model, and last the \DEF model.
With the \RISK and the \RRT model the difference was smaller for the second order group, \ie,
where the differential privacy explanation was shown first.

All requirements for conducting a between-subject two-way ANOVA were met.
In particular, D'Agostino's K-squared tests for normality did not reveal any significant deviation of the differences from a normal distribution for any of the three conditions.
Also, Levene's test indicated equality of variances between the conditions.
The ANOVA did not reveal any significant effect of the explanatory model on the differences in objective comprehension ($F \approx1.245$, $p\approx0.293$).
There was no indication of a significant effect of the order of the explanations or a significant interaction of the explanatory model and the order.
\Cref{fig:boxplot-objective} shows that the interquartile range of the differences was lower with \RISK and \RRT than with \DEF.
For these reasons, we conducted post-hoc one-tailed t-tests to investigate
whether the mean differences with the \RISK model and the \RRT model for differential privacy protection differed significantly from the \DEF model.
The results indicated that the mean difference was smaller with the \RISK model than with the \DEF model ($t \approx-1.008$, $p\approx0.159$, Cohens $d\approx0.262$).
There was a tendency for the mean difference to be smaller with the \RRT model than with the \DEF model ($t\approx-1.467$, $p\approx0.074$).

These results support (H2), suggesting that the users' objective comprehension
concerning the privacy protection provided by
differential privacy is enhanced through the \RISK model and may be enhanced through the \RRT model.

\subsection{Comparison of $k$-Anonymity and Differential Privacy}
\begin{figure}[tb]
	\begin{subfigure}[t]{.45\columnwidth}
		\centering
		\includegraphics[width=\columnwidth]{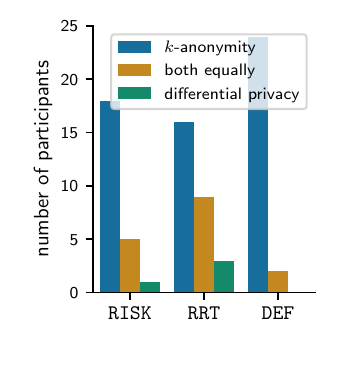}
		\caption{Comprehensible}
		\label{fig:comparisionComprehensible}
	\end{subfigure}%
	\begin{subfigure}[t]{.45\columnwidth}
		\centering
		\includegraphics[width=\columnwidth]{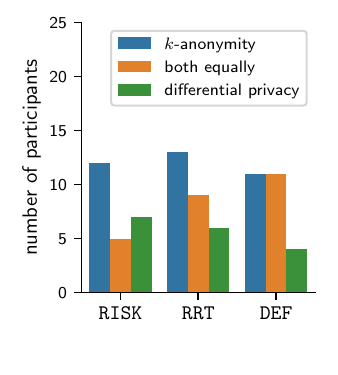}
		\caption{Prevent}
		\label{fig:comparisionPrevent}
	\end{subfigure}
	\caption{Comparison regarding comprehensibility and privacy prevention.}
	\label{fig:comparison}
	\vspace{-1em}
\end{figure}

We present the answers to the direct comparison between the differential privacy mechanism and the $k$-anonymity mechanism in \Cref{fig:comparison}.
Few participants rated the level of privacy protection and thus the implications of differential privacy as more comprehensible than that of $k$-anonymity (see \Cref{fig:comparisionComprehensible}).
In \DEF, nobody rated the differential privacy explanation more
comprehensible than $k$-anonymity.
These results support (H1), which states that the privacy protection of $k$-anonymity
is easier to comprehend than the privacy protection of differential privacy.

The overall answer about which mechanism was better at preventing a data breach was in favor of $k$-anonymity with the \RISK model and the \RRT model
(cf. \Cref{fig:comparisionPrevent}).
In \DEF, $11$ participants said that both were equally good
at preventing a data breach.

\subsection{Effects of Level of Education and Numeracy Skills}
\begin{figure}[tb]
	\begin{subfigure}[t]{.45\columnwidth}
		\centering
		\includegraphics[width=\columnwidth]{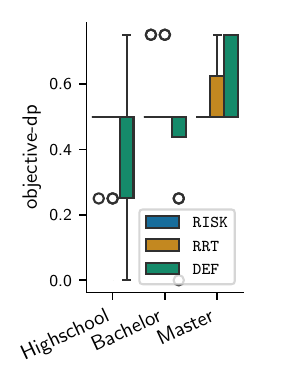}
		\caption{Education level}
		\label{fig:correlationEdu}
	\end{subfigure}%
	\begin{subfigure}[t]{.45\columnwidth}
		\centering
		\includegraphics[width=\columnwidth]{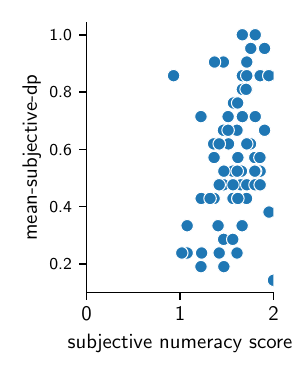}
		\caption{Subjective numeracy}
		\label{fig:correlationNumeracy}
	\end{subfigure}
	\caption{Correlations of comprehension.}
	\label{fig:correlation}
\end{figure}

In \Cref{fig:correlationEdu}, we show that objective comprehension scores increase with
higher levels of education across all explanatory models, with $r\approx0.285$,
and $p\approx0.008$.
Especially for the \DEF model, a higher level of education is associated with a
higher objective comprehension score.
Thus, we can confirm~(H3), which states that a high level of education helps
users to comprehend the privacy protection of differential privacy.

Our participants' objective and subjective numeracy skills were high overall; we had no participants with a score below~$0.9$ (see \Cref{fig:correlationNumeracy}).
The participants' subjective numeracy skills also correlated positively
with their subjective comprehension scores for differential privacy protection, with $r\approx0.299$, $p\approx0.008$.
We did not find any correlation between the objective numeracy skills and objective comprehension score.
Regarding the subjective comprehension, we can confirm (H3).
We did not find any other correlations between the objective or subjective comprehension scores and the level of privacy concerns.

\subsection{Exclusion of Knowledgeable Participants}
Prior knowledge of privacy definitions may have influenced the study results\footnote{We could not exclude the possibility of an overlap of participants who took part in the pilot study and the main study.
	However, by excluding participants with prior knowledge of privacy,
	we controlled for overlap and thus verified that the main results would remain valid.}.
Therefore, we reran our analysis, excluding all participants who indicated that they were
already aware of one or more privacy mechanisms, particularly differential privacy.
This resulted in only $56$ participants who completed the study, so the results are limited in their power.
However, most findings remained unchanged after these adaptions.

The privacy protection provided by $k$-anonymity was subjectively rated as significantly more easily understood than the privacy protection provided by differential privacy among respondents in the \RISK
and in the \DEF group,
but not in the \RRT group.

The mean difference between the subjective comprehension scores of
$k$-anonymity and differential privacy was again smallest with the \RRT model, followed by the \RISK model.
There was still a significant effect of the explanatory model on the differences in the subjective comprehension scores ($F \approx8.814$, $p<0.001$,~$\eta^{2}\approx0.2570$) and no significant effect from the order of the
explanations or from the interaction between the explanatory model and the order of the explanations.
One-tailed t-tests indicated that the mean difference between the subjective comprehension scores for the privacy protection provided by $k$-anonymity and differential privacy was significantly smaller when using the
\RRT model
compared to the \DEF model.

The scores regarding the objective comprehension of differential privacy protection were highest with the \RISK, followed by the \RRT model. However, the mean difference between the privacy protection of $k$-anonymity and of differential privacy was now smallest with the \RRT model.
There was still no significant
effect on differences in the objective comprehension scores.

The findings regarding the correlations were similar to the findings when knowledgeable participants were included.
There was again a significant positive correlation between subjective comprehension scores and subjective numeracy skills ($r\approx0.349$,~$p\approx0.008$).
The level of education also indicated a positive direction to the objective
comprehension ($r\approx0.251$, $p\approx0.067$).
Remarkably, we found a significant positive correlation between the objective
and subjective comprehension for $k$-anonymity ($r\approx0.361$, $p\approx0.006$).
Again, we did not find any other expected correlations.

\subsection{Summary of Findings}
The following points represent our main findings:
\begin{itemize}
	\item The privacy protection of $k$-anonymity was only rated as
	significantly more easily understood subjectively than the privacy
	protection of differential privacy with the \RISK and the \DEF models.
	\item The privacy protection of $k$-anonymity is objectively easier to
	comprehend than the privacy protection of differential privacy
	(independent of the explanatory model).
	\item The \RISK and \RRT models significantly enhanced the subjective comprehensibility of differential privacy protection to greater extent than the \DEF model.
	\item The objective comprehension of differential privacy protection is enhanced by the \RISK model and may be enhanced through the \RRT model.
	\item We find a positive correlation between the level of education and
	objective comprehension.
	\item Participants with high subjective numeracy skills had also high subjective comprehension scores. Moreover, participants with high objective numeracy skills had also higher objective comprehension scores.	
\end{itemize}

\section{Discussion}
\label{sec:discussion}
Our study was motivated by investigating various models attempting to explain
privacy protection of differential privacy.
We compared the comprehensibility of three different explanatory models with the privacy protection of $k$-anonymity as a baseline for comprehensibility.
In the following section, we put our results into a larger context.
One especially salient aspect is the suitability of our explanatory models for
understanding differential privacy protection.
We also reflect on the differences in how users understand privacy protection of
differential privacy compared to the protection of $k$-anonymity.
Finally, we present our thoughts on the influence of the level of education
and numeracy skills. 


\subsubsection*{The different explanatory models support the comprehension of privacy protection to varying extents}
The subjective comprehension is
significantly enhanced by the \RRT and \RISK models of explanation.
The \DEF model was the most difficult to understand with respect to the privacy protection.
This result fits well with~\cite{cummings} and confirms that
the pure definition of differential privacy is not that easy to understand.
Nevertheless, the \RRT model helped people to comprehend differential privacy protection significantly better than the \RISK model did.
We therefore suggest that the \RRT model, as a metaphor, is easier to imagine.
However, Karegar et~al.~\cite{KaregarAF22} advised being careful with metaphors,
as users find them difficult to transfer to other contexts.
We believe that a transfer to other contexts is easier with our explanatory \RRT model
because we describe privacy protection and not the mechanism itself.

\subsubsection*{Explanatory models do not contribute to objective comprehension}
Whereas subjective comprehension was enhanced,
we found no difference on objective comprehension of differential privacy
protection.
Surprisingly, our objective comprehension scores for all differential privacy
models were low, with a mean of $\approx0.5$.
This may be either due to the complexity of our explanatory models
or to the difficulty of our objective questions.
In order to compare $k$-anonymity and differential privacy,
we aligned the questions, but
we compared two different mechanisms.
Our questions on objective comprehension were therefore unable to fully capture the
subtleties of differential privacy and thus of our explanatory model.
Future studies should therefore adapt the questions accordingly to better
capture objective comprehension.

\subsubsection*{The privacy protection of $k$-anonymity is more comprehensible than that by differentiated privacy}
We can summarize by stating that
the privacy protection provided by $k$-anonymity seems to be
subjectively easier to comprehend than that of differential privacy.
Differential privacy protection explained with the \RRT model seems
almost as easy to comprehend as the privacy protection provided by $k$-anonymity.
Nevertheless, even here more than half of the participants rated
$k$-anonymity as subjectively more comprehensible when asked to directly compare the comprehension of the privacy protection of both mechanisms.
This result is in line with the findings of Valdez~et~al.~\cite{ValdezZ19},
who evaluated users' willingness to share personal health data when
applying privacy-preserving techniques such as $k$-anonymity or differential privacy.
The perception of privacy was rated more strongly for $k$-anonymity than
for differential privacy.
The authors assumed that this was because the protection of differential privacy
was too difficult to conceptualize.
Our results show that the \RRT model enhances the subjective comprehensibility
significantly compared to the other two explanatory models, and
also yields the best scores for the participants' subjective comprehension of
the privacy protection in comparison to the privacy protection of $k$-anonymity.
We therefore argue that this model can serve as a basis for further studies.

In addition, we have high scores concerning objective comprehension for $k$-anonymity.
This is probably due to the fact that the privacy parameter $k$ has a direct implication for data protection,
which can be understood independently of the data~\cite{CliftonT13}:
the parameter is related to the legal concept of individual identifiability. 
Our explanatory models help establish this relationship to identifiability,
but the implications must be explained in terms of a use case.
With $k$-anonymity, privacy protection and the mechanism itself
can be easily visualized with a data set.
Hence, a non-expert can verifiy that the published data set is indeed $k$-anonymous~\cite{FriedmanWS08}.
When communicating differential privacy guarantees,
past studies have either attempted to explain or visualize the mechanism itself~ (\cite{ecostrategy,theorybased})
or the risk associated with an $\varepsilon$ (\cite{leeclifton,franzen}).
Similar to research by Nanayakkara et al.~\cite{NanayakkaraB0HR22},
future research should find explanations that convey both the risks and the
privacy-utility tradeoff.
We believe that the implications are better understood if the explanation is not use-case dependent.

\subsubsection*{Different levels of education need different explanations}
We observed that levels of education and numeracy skills led to better objective
and subjective comprehension.
Differential privacy provides a quantitative mathematical definition,
so it makes sense that
numeracy skills would be helpful in understanding the privacy protection
of differential privacy.
However, we also observed a positive correlation between the comprehension
and subjective numeracy skills.
These results do not correspond to the Dunning-Kruger effect~\cite{kruger1999unskilled}.
However, this might be due to the fact that our sample was very homogeneous and highly educated.
Nevertheless, different target groups should receive different explanations.
Our target group was end users;
further research is needed to determine whether our explanatory models can
help other audiences, \eg developers, choose a suitable ~$\varepsilon$.


\section{Related Work}
\label{sec:relatedwork}
Considering the demand for strong privacy guarantees,
there is a plethora of work on developing new algorithms with
differential privacy guarantees that
focus on improving the privacy-utility tradeoff
\cite{dworkEpsilon,Patel2018a,WangZFY20}.
A smaller $\varepsilon$ leads to stronger privacy.
However, the problem of how exactly to determine the value of $\varepsilon$
remains a major challenge.
According to Dwork, the value of~$\varepsilon$ is a \enquote{social decision} \cite{dworkEpsilon}.
Therefore, various approaches have been proposed to quantify the privacy guarantees by translating $\varepsilon$ into a privacy risk
~\cite{leeclifton,MehnerVT21} or by using metaphors to describe the differential privacy mechanism~\cite{bullek,KaregarAF22,SmartSV22}.

The privacy risk as well as the randomized response technique may serve as
explanatory models for the differential privacy guarantee~\cite{bullek,SmartSV22}.
The \RRT model has been researched with regard to users' trust and
comprehension~\cite{bullek,trustunderstanding} of the technique.
Bullek et al.~\cite{bullek} focused on describing the randomized response technique by using a spinner metaphor.
Smart et al.~\cite{SmartSV22} investigated explanations of a differential privacy mechanism,
hiding the $\varepsilon$ used and evaluating users' willingness to share data.
However, it has not yet been determined whether users' understanding of this
technique is sufficient.
Franzen et al.~\cite{franzen} evaluated the \RISK approach empirically, with
a focus on how to communicate this risk.
Our present study, is a first step in evaluating
comprehensibility of the \RISK model as well as of the \RRT model.
Furthermore, the randomized response technique has generally been researched in
isolation;
in contrast, this study evaluates the technique as
means of explaining differential privacy in general.

Some previous studies have evaluated descriptions for differential privacy.
The work of Cummings et al.~\cite{cummings} took a look at users' expectations
that arose from descriptions of differential privacy mechanisms already in the industry.
According to the authors of that study, existing descriptions fail to explain the privacy
guarantee provided by differential privacy in that users' expectations are set
arbitrarily.
We have aimed for an explanation that would improve users' comprehension of the
differential privacy guarantee.
Instead of looking at previously written descriptions,
we have evaluated explanatory models that are intended to facilitate users' comprehension.

Karegar et al.~\cite{KaregarAF22} used blurred images as a metaphor to explain the privacy protection provided by differential privacy.
The authors noted that the explanations help to communicate the fact that introduced noise protects privacy and that there is some privacy-utility tradeoff.
However, their explanations did not directly imply privacy protection of a particular $\varepsilon$.
ViP~\cite{NanayakkaraB0HR22} is a tool for supporting the decision of setting/splitting $\varepsilon$ across queries.
Nanayakkara et~al.~\cite{NanayakkaraB0HR22} used \RISK~(\cite{leeclifton})
to show the risk for a different number of users and set $\varepsilon$.
The authors then depicted the privacy-utility tradeoff for data analysts.
In our study, we have focused on end users as the target group of such explanations.

Xiong et al.~\cite{xiong} researched how different explanations for
differential privacy influenced users' willingness to share their personal data.
The authors found that explanations focusing on the implications of differential
privacy instead of the technical aspects increased users' understanding and
willingness to share personal data.
Nanayakkara et al.~\cite{NanayakkaraSCKR23} used odds-based explanations based on the \RISK model inspired by \cite{leeclifton}.
They then compared their explanations to the explanations provided by Xiong et al.~\cite{xiong}.
Our study takes a further step towards a more comprehensive explanation of
differential privacy, in that we evaluate different explanatory models
that all focus on the implications of the mechanism,
\ie, the privacy guarantee provided.

Our experimental setup, comparing users' comprehension provided by $k$-anonymity to that provided by differential privacy was inspired by
Valdez et al.~\cite{ValdezZ19}.
They examined how understandings of privacy concern
change depending on what data is collected and how it is used.
The degree of privacy of $k$-anonymity was also given with
\enquote{indistinguishability}~\cite{ValdezZ19}.
To explain differential privacy, the concept of \enquote{exceptionality} was used,
which indicated how exceptional one is among all other respondents.
The results of their study suggest that being part of a larger crowd
($k$-anonymity) appears to be more privacy protective than differential privacy does.
The authors hypothesized that this is due to the explanation of exceptionality.
With our work, we have provided models for
explaining differential privacy protection
and have evaluated them for their comprehensibility.
Due to our comparison with users' comprehension of the privacy protection
of $k$-anonymity, we have been able to help shed light on the extent of users' comprehension.

\section{Conclusion}
\label{sec:conclusion}
We can conclude that different explanatory models indeed help people to comprehend the privacy protection provided by differential privacy.
Our results confirm that the \RISK and \RRT model enhance users' subjective
comprehension provided by differential privacy protection better than the \DEF model does.
We have
therefore presented a way to effectively explain the privacy protection of a
Laplacian differential privacy mechanism.
Moreover, the privacy protection provided by $k$-anonymity was more comprehensible
than that provided by differential privacy.
The \RRT model yields the best scores for the
participants' subjective comprehension.
Therefore, we can conclude that \RRT can serve as a basis for further~studies.

\begin{acks}
This work is partially funded by the European Union (NextGenerationEU).
It is also supported by the German Federal Ministry of Education and Research~(BMBF)
as part of the research projects FreeMove and GANGES under reference number
01UV2090B and 16KISA034, respectively.
\end{acks}



\begin{thebibliography}{31}


\ifx \showCODEN    \undefined \def \showCODEN     #1{\unskip}     \fi
\ifx \showDOI      \undefined \def \showDOI       #1{#1}\fi
\ifx \showISBNx    \undefined \def \showISBNx     #1{\unskip}     \fi
\ifx \showISBNxiii \undefined \def \showISBNxiii  #1{\unskip}     \fi
\ifx \showISSN     \undefined \def \showISSN      #1{\unskip}     \fi
\ifx \showLCCN     \undefined \def \showLCCN      #1{\unskip}     \fi
\ifx \shownote     \undefined \def \shownote      #1{#1}          \fi
\ifx \showarticletitle \undefined \def \showarticletitle #1{#1}   \fi
\ifx \showURL      \undefined \def \showURL       {\relax}        \fi
\providecommand\bibfield[2]{#2}
\providecommand\bibinfo[2]{#2}
\providecommand\natexlab[1]{#1}
\providecommand\showeprint[2][]{arXiv:#2}

\bibitem[Bullek et~al\mbox{.}(2017)]%
        {bullek}
\bibfield{author}{\bibinfo{person}{Brooke Bullek}, \bibinfo{person}{Stephanie
  Garboski}, \bibinfo{person}{Darakhshan~J. Mir}, {and}
  \bibinfo{person}{Evan~M. Peck}.} \bibinfo{year}{2017}\natexlab{}.
\newblock \showarticletitle{Towards Understanding Differential Privacy: When Do
  People Trust Randomized Response Technique?}. In
  \bibinfo{booktitle}{\emph{CHI~'17: Proceedings of the 2017 Conference on
  Human Factors in Computing Systems}}. \bibinfo{publisher}{ACM},
  \bibinfo{pages}{3833--3837}.
\newblock
\urldef\tempurl%
\url{https://doi.org/10.1145/3025453.3025698}
\showDOI{\tempurl}


\bibitem[Clifton and Tassa(2013)]%
        {CliftonT13}
\bibfield{author}{\bibinfo{person}{Chris Clifton} {and} \bibinfo{person}{Tamir
  Tassa}.} \bibinfo{year}{2013}\natexlab{}.
\newblock \showarticletitle{On syntactic anonymity and differential privacy}.
  In \bibinfo{booktitle}{\emph{ICDEW~'13: IEEE 29th International Conference on
  Data Engineering Workshops}}. \bibinfo{pages}{88--93}.
\newblock
\urldef\tempurl%
\url{https://doi.org/10.1109/ICDEW.2013.6547433}
\showDOI{\tempurl}


\bibitem[Cummings et~al\mbox{.}(2021)]%
        {cummings}
\bibfield{author}{\bibinfo{person}{Rachel Cummings}, \bibinfo{person}{Gabriel
  Kaptchuk}, {and} \bibinfo{person}{Elissa~M. Redmiles}.}
  \bibinfo{year}{2021}\natexlab{}.
\newblock \showarticletitle{"I need a better description": An Investigation
  Into User Expectations For Differential Privacy}. In
  \bibinfo{booktitle}{\emph{{CCS} '21: Proceedings of the 2021 {ACM} {SIGSAC}
  Conference on Computer and Communications Security, Virtual Event, Republic
  of Korea, November 15 - 19, 2021}}. \bibinfo{publisher}{{ACM}},
  \bibinfo{pages}{3037--3052}.
\newblock
\urldef\tempurl%
\url{https://doi.org/10.1145/3460120.3485252}
\showDOI{\tempurl}


\bibitem[Dwork(2006)]%
        {diffpri}
\bibfield{author}{\bibinfo{person}{Cynthia Dwork}.}
  \bibinfo{year}{2006}\natexlab{}.
\newblock \showarticletitle{Differential Privacy}. In
  \bibinfo{booktitle}{\emph{ICALP~'06: Automata, Languages and Programming,
  33rd International Colloquium, Proceedings, Part {II}}}
  \emph{(\bibinfo{series}{Lecture Notes in Computer Science},
  Vol.~\bibinfo{volume}{4052})}. \bibinfo{publisher}{Springer},
  \bibinfo{pages}{1--12}.
\newblock
\urldef\tempurl%
\url{https://doi.org/10.1007/11787006\_1}
\showDOI{\tempurl}


\bibitem[Dwork(2008)]%
        {dworkEpsilon}
\bibfield{author}{\bibinfo{person}{Cynthia Dwork}.}
  \bibinfo{year}{2008}\natexlab{}.
\newblock \showarticletitle{Differential Privacy: {A} Survey of Results}. In
  \bibinfo{booktitle}{\emph{{TAMC}~'08: Theory and Applications of Models of
  Computation, 5th International Conference}} \emph{(\bibinfo{series}{Lecture
  Notes in Computer Science}, Vol.~\bibinfo{volume}{4978})}.
  \bibinfo{publisher}{Springer}, \bibinfo{pages}{1--19}.
\newblock
\urldef\tempurl%
\url{https://doi.org/10.1007/978-3-540-79228-4\_1}
\showDOI{\tempurl}


\bibitem[Fagerlin et~al\mbox{.}(7 09)]%
        {numeracyFagerlin}
\bibfield{author}{\bibinfo{person}{Angela Fagerlin}, \bibinfo{person}{Brian
  Zikmund-Fisher}, \bibinfo{person}{Peter Ubel}, \bibinfo{person}{Aleksandra
  Jankovic}, \bibinfo{person}{Holly Derry}, {and} \bibinfo{person}{Dylan
  Smith}.} \bibinfo{year}{2007-09}\natexlab{}.
\newblock \showarticletitle{Measuring Numeracy Without a Math Test: Development
  of the Subjective Numeracy Scale}.
\newblock \bibinfo{journal}{\emph{Medical decision making : an international
  journal of the Society for Medical Decision Making}}  \bibinfo{volume}{27}
  (\bibinfo{year}{2007-09}), \bibinfo{pages}{672--80}.
\newblock
\urldef\tempurl%
\url{https://doi.org/10.1177/0272989X07304449}
\showDOI{\tempurl}


\bibitem[Franzen et~al\mbox{.}(2022)]%
        {franzen}
\bibfield{author}{\bibinfo{person}{Daniel Franzen},
  \bibinfo{person}{Saskia~Nu{\~{n}}ez von Voigt}, \bibinfo{person}{Peter
  S{\"{o}}rries}, \bibinfo{person}{Florian Tschorsch}, {and}
  \bibinfo{person}{Claudia M{\"{u}}ller{-}Birn}.}
  \bibinfo{year}{2022}\natexlab{}.
\newblock \showarticletitle{Am {I} Private and If So, how Many?: Communicating
  Privacy Guarantees of Differential Privacy with Risk Communication Formats}.
  In \bibinfo{booktitle}{\emph{CCS~'22: Proceedings of the 2022 {ACM} {SIGSAC}
  Conference on Computer and Communications Security, Los Angeles, CA, USA,
  November 7-11, 2022}}. \bibinfo{publisher}{{ACM}},
  \bibinfo{pages}{1125--1139}.
\newblock
\urldef\tempurl%
\url{https://doi.org/10.1145/3548606.3560693}
\showDOI{\tempurl}


\bibitem[Friedman et~al\mbox{.}(2008)]%
        {FriedmanWS08}
\bibfield{author}{\bibinfo{person}{Arik Friedman}, \bibinfo{person}{Ran Wolff},
  {and} \bibinfo{person}{Assaf Schuster}.} \bibinfo{year}{2008}\natexlab{}.
\newblock \showarticletitle{Providing \emph{k}-anonymity in data mining}.
\newblock \bibinfo{journal}{\emph{The {VLDB} Journal}} \bibinfo{volume}{17},
  \bibinfo{number}{4} (\bibinfo{year}{2008}), \bibinfo{pages}{789--804}.
\newblock
\urldef\tempurl%
\url{https://doi.org/10.1007/S00778-006-0039-5}
\showDOI{\tempurl}


\bibitem[Fung et~al\mbox{.}(2010)]%
        {FungWCY10}
\bibfield{author}{\bibinfo{person}{Benjamin C.~M. Fung}, \bibinfo{person}{Ke
  Wang}, \bibinfo{person}{Rui Chen}, {and} \bibinfo{person}{Philip~S. Yu}.}
  \bibinfo{year}{2010}\natexlab{}.
\newblock \showarticletitle{Privacy-preserving data publishing: {A} survey of
  recent developments}.
\newblock \bibinfo{journal}{\emph{Comput. Surveys}} \bibinfo{volume}{42},
  \bibinfo{number}{4} (\bibinfo{year}{2010}), \bibinfo{pages}{14:1--14:53}.
\newblock
\urldef\tempurl%
\url{https://doi.org/10.1145/1749603.1749605}
\showDOI{\tempurl}


\bibitem[Hsu et~al\mbox{.}(2014)]%
        {ecostrategy}
\bibfield{author}{\bibinfo{person}{Justin Hsu}, \bibinfo{person}{Marco
  Gaboardi}, \bibinfo{person}{Andreas Haeberlen}, \bibinfo{person}{Sanjeev
  Khanna}, \bibinfo{person}{Arjun Narayan}, \bibinfo{person}{Benjamin~C.
  Pierce}, {and} \bibinfo{person}{Aaron Roth}.}
  \bibinfo{year}{2014}\natexlab{}.
\newblock \showarticletitle{Differential Privacy: An Economic Method for
  Choosing Epsilon}. In \bibinfo{booktitle}{\emph{CSF~'14: {IEEE} 27th Computer
  Security Foundations Symposium}}. \bibinfo{publisher}{{IEEE} Computer
  Society}, \bibinfo{pages}{398--410}.
\newblock
\urldef\tempurl%
\url{https://doi.org/10.1109/CSF.2014.35}
\showDOI{\tempurl}


\bibitem[Karegar et~al\mbox{.}(2022)]%
        {KaregarAF22}
\bibfield{author}{\bibinfo{person}{Farzaneh Karegar},
  \bibinfo{person}{Ala~Sarah Alaqra}, {and} \bibinfo{person}{Simone
  Fischer{-}H{\"{u}}bner}.} \bibinfo{year}{2022}\natexlab{}.
\newblock \showarticletitle{Exploring User-Suitable Metaphors for
  Differentially Private Data Analyses}. In
  \bibinfo{booktitle}{\emph{SOUPS~'22: Proceedings of the Eighteenth Symposium
  on Usable Privacy and Security, Boston, MA, USA, August 7-9, 2022}}.
  \bibinfo{publisher}{{USENIX} Association}, \bibinfo{pages}{175--193}.
\newblock
\urldef\tempurl%
\url{https://www.usenix.org/conference/soups2022/presentation/karegar}
\showURL{%
\tempurl}


\bibitem[Keller and Siegrist(2009)]%
        {Keller09}
\bibfield{author}{\bibinfo{person}{Carmen Keller} {and}
  \bibinfo{person}{Michael Siegrist}.} \bibinfo{year}{2009}\natexlab{}.
\newblock \showarticletitle{Effect of Risk Communication Formats on Risk
  Perception Depending on Numeracy}.
\newblock \bibinfo{journal}{\emph{Medical Decision Making}}
  \bibinfo{volume}{29}, \bibinfo{number}{4} (\bibinfo{year}{2009}),
  \bibinfo{pages}{483--490}.
\newblock
\urldef\tempurl%
\url{https://doi.org/10.1177/0272989X09333122}
\showDOI{\tempurl}


\bibitem[Kruger and Dunning(1999)]%
        {kruger1999unskilled}
\bibfield{author}{\bibinfo{person}{Justin Kruger} {and} \bibinfo{person}{David
  Dunning}.} \bibinfo{year}{1999}\natexlab{}.
\newblock \showarticletitle{Unskilled and unaware of it: how difficulties in
  recognizing one's own incompetence lead to inflated self-assessments.}
\newblock \bibinfo{journal}{\emph{Journal of personality and social
  psychology}} \bibinfo{volume}{77}, \bibinfo{number}{6}
  (\bibinfo{year}{1999}), \bibinfo{pages}{1121}.
\newblock


\bibitem[Landsheer et~al\mbox{.}(1999)]%
        {trustunderstanding}
\bibfield{author}{\bibinfo{person}{Johannes~A Landsheer},
  \bibinfo{person}{Peter Van Der~Heijden}, {and} \bibinfo{person}{Ger
  Van~Gils}.} \bibinfo{year}{1999}\natexlab{}.
\newblock \showarticletitle{Trust and understanding, two psychological aspects
  of randomized response}.
\newblock \bibinfo{journal}{\emph{Quality and Quantity}} \bibinfo{volume}{33},
  \bibinfo{number}{1} (\bibinfo{year}{1999}), \bibinfo{pages}{1--12}.
\newblock
\urldef\tempurl%
\url{https://doi.org/10.1023/A:1004361819974}
\showDOI{\tempurl}


\bibitem[Lee and Clifton(2011)]%
        {leeclifton}
\bibfield{author}{\bibinfo{person}{Jaewoo Lee} {and} \bibinfo{person}{Chris
  Clifton}.} \bibinfo{year}{2011}\natexlab{}.
\newblock \showarticletitle{How Much Is Enough? Choosing $\epsilon$ for
  Differential Privacy}. In \bibinfo{booktitle}{\emph{ISC~'11: Information
  Security, 14th International Conference}}. \bibinfo{publisher}{Springer},
  \bibinfo{pages}{325--340}.
\newblock
\urldef\tempurl%
\url{https://doi.org/10.1007/978-3-642-24861-0_22}
\showDOI{\tempurl}


\bibitem[Li et~al\mbox{.}(2007)]%
        {LiLV07}
\bibfield{author}{\bibinfo{person}{Ninghui Li}, \bibinfo{person}{Tiancheng Li},
  {and} \bibinfo{person}{Suresh Venkatasubramanian}.}
  \bibinfo{year}{2007}\natexlab{}.
\newblock \showarticletitle{t-Closeness: Privacy Beyond k-Anonymity and
  l-Diversity}. In \bibinfo{booktitle}{\emph{ICDE~'07: Proceedings of the 23rd
  International Conference on Data Engineering}}. \bibinfo{publisher}{{IEEE}
  Computer Society}, \bibinfo{pages}{106--115}.
\newblock
\urldef\tempurl%
\url{https://doi.org/10.1109/ICDE.2007.367856}
\showDOI{\tempurl}


\bibitem[Lipkus et~al\mbox{.}(1 02)]%
        {numeracyLipkus}
\bibfield{author}{\bibinfo{person}{Isaac Lipkus}, \bibinfo{person}{Greg Samsa},
  {and} \bibinfo{person}{Barbara Rimer}.} \bibinfo{year}{2001-02}\natexlab{}.
\newblock \showarticletitle{General Performance on a Numeracy Scale Among
  Highly Educated Samples}.
\newblock \bibinfo{journal}{\emph{Medical decision making : an international
  journal of the Society for Medical Decision Making}}  \bibinfo{volume}{21}
  (\bibinfo{year}{2001-02}), \bibinfo{pages}{37--44}.
\newblock
\urldef\tempurl%
\url{https://doi.org/10.1177/0272989X0102100105}
\showDOI{\tempurl}


\bibitem[Machanavajjhala et~al\mbox{.}(2006)]%
        {MachanavajjhalaGKV06}
\bibfield{author}{\bibinfo{person}{Ashwin Machanavajjhala},
  \bibinfo{person}{Johannes Gehrke}, \bibinfo{person}{Daniel Kifer}, {and}
  \bibinfo{person}{Muthuramakrishnan Venkitasubramaniam}.}
  \bibinfo{year}{2006}\natexlab{}.
\newblock \showarticletitle{l-Diversity: Privacy Beyond k-Anonymity}. In
  \bibinfo{booktitle}{\emph{ICDE~'06: Proceedings of the 22nd International
  Conference on Data Engineering}}. \bibinfo{publisher}{{IEEE} Computer
  Society}, \bibinfo{pages}{24}.
\newblock
\urldef\tempurl%
\url{https://doi.org/10.1109/ICDE.2006.1}
\showDOI{\tempurl}


\bibitem[Malhotra et~al\mbox{.}(2004)]%
        {MalhotraKA04}
\bibfield{author}{\bibinfo{person}{Naresh~K. Malhotra},
  \bibinfo{person}{Sung~S. Kim}, {and} \bibinfo{person}{James Agarwal}.}
  \bibinfo{year}{2004}\natexlab{}.
\newblock \showarticletitle{Internet Users' Information Privacy Concerns
  {(IUIPC):} The Construct, the Scale, and a Causal Model}.
\newblock \bibinfo{journal}{\emph{Information Systems Research}}
  \bibinfo{volume}{15}, \bibinfo{number}{4} (\bibinfo{year}{2004}),
  \bibinfo{pages}{336--355}.
\newblock
\urldef\tempurl%
\url{https://doi.org/10.1287/isre.1040.0032}
\showDOI{\tempurl}


\bibitem[Mehner et~al\mbox{.}(2021)]%
        {MehnerVT21}
\bibfield{author}{\bibinfo{person}{Luise Mehner},
  \bibinfo{person}{Saskia~Nu{\~{n}}ez von Voigt}, {and}
  \bibinfo{person}{Florian Tschorsch}.} \bibinfo{year}{2021}\natexlab{}.
\newblock \showarticletitle{Towards Explaining Epsilon: {A} Worst-Case Study of
  Differential Privacy Risks}. In \bibinfo{booktitle}{\emph{EuroS{\&}P~'21:
  {IEEE} European Symposium on Security and Privacy Workshops, Vienna, Austria,
  September 6-10, 2021}}. \bibinfo{publisher}{{IEEE}},
  \bibinfo{pages}{328--331}.
\newblock
\urldef\tempurl%
\url{https://doi.org/10.1109/EUROSPW54576.2021.00041}
\showDOI{\tempurl}


\bibitem[Naldi and D'Acquisto(2015)]%
        {theorybased}
\bibfield{author}{\bibinfo{person}{Maurizio Naldi} {and}
  \bibinfo{person}{Giuseppe D'Acquisto}.} \bibinfo{year}{2015}\natexlab{}.
\newblock \showarticletitle{Differential Privacy: An Estimation Theory-Based
  Method for Choosing Epsilon}.
\newblock \bibinfo{journal}{\emph{arXiv preprint}}
  \bibinfo{volume}{abs/1510.00917} (\bibinfo{year}{2015}).
\newblock


\bibitem[Nanayakkara et~al\mbox{.}(2022)]%
        {NanayakkaraB0HR22}
\bibfield{author}{\bibinfo{person}{Priyanka Nanayakkara},
  \bibinfo{person}{Johes Bater}, \bibinfo{person}{Xi He},
  \bibinfo{person}{Jessica Hullman}, {and} \bibinfo{person}{Jennie Rogers}.}
  \bibinfo{year}{2022}\natexlab{}.
\newblock \showarticletitle{Visualizing Privacy-Utility Trade-Offs in
  Differentially Private Data Releases}.
\newblock \bibinfo{journal}{\emph{Proceedings on Privacy Enhancing
  Technologies}} \bibinfo{volume}{2022}, \bibinfo{number}{2}
  (\bibinfo{year}{2022}), \bibinfo{pages}{601--618}.
\newblock
\urldef\tempurl%
\url{https://doi.org/10.2478/popets-2022-0058}
\showDOI{\tempurl}


\bibitem[Nanayakkara et~al\mbox{.}(2023)]%
        {NanayakkaraSCKR23}
\bibfield{author}{\bibinfo{person}{Priyanka Nanayakkara},
  \bibinfo{person}{Mary~Anne Smart}, \bibinfo{person}{Rachel Cummings},
  \bibinfo{person}{Gabriel Kaptchuk}, {and} \bibinfo{person}{Elissa~M.
  Redmiles}.} \bibinfo{year}{2023}\natexlab{}.
\newblock \showarticletitle{What Are the Chances? Explaining the Epsilon
  Parameter in Differential Privacy}. In \bibinfo{booktitle}{\emph{32nd
  {USENIX} Security Symposium, {USENIX} Security 2023, Anaheim, CA, USA, August
  9-11, 2023}}. \bibinfo{publisher}{{USENIX} Association}.
\newblock
\urldef\tempurl%
\url{https://www.usenix.org/conference/usenixsecurity23/presentation/nanayakkara}
\showURL{%
\tempurl}


\bibitem[{Patel} and {Jethava}(2018)]%
        {Patel2018a}
\bibfield{author}{\bibinfo{person}{K. {Patel}} {and} \bibinfo{person}{G.~B.
  {Jethava}}.} \bibinfo{year}{2018}\natexlab{}.
\newblock \showarticletitle{Privacy Preserving Techniques for Big Data: A
  Survey}. In \bibinfo{booktitle}{\emph{ICICCT~'18: Proceedings of the 2018
  Second International Conference on Inventive Communication and Computational
  Technologies}}. \bibinfo{pages}{194--199}.
\newblock
\urldef\tempurl%
\url{https://doi.org/10.1109/ICICCT.2018.8473289}
\showDOI{\tempurl}


\bibitem[Schrager(2018)]%
        {numeracySchrager}
\bibfield{author}{\bibinfo{person}{Sarina~B. Schrager}.}
  \bibinfo{year}{2018}\natexlab{}.
\newblock \showarticletitle{Five Ways to Communicate Risks So That Patients
  Understand.}
\newblock \bibinfo{journal}{\emph{Family practice management}}
  \bibinfo{volume}{25 6} (\bibinfo{year}{2018}), \bibinfo{pages}{28--31}.
\newblock


\bibitem[Smart et~al\mbox{.}({[n.\,d.]})]%
        {SmartSV22}
\bibfield{author}{\bibinfo{person}{Mary~Anne Smart}, \bibinfo{person}{Dhruv
  Sood}, {and} \bibinfo{person}{Kristen Vaccaro}.}
  \bibinfo{year}{[n.\,d.]}\natexlab{}.
\newblock \showarticletitle{Understanding Risks of Privacy Theater with
  Differential Privacy}.
\newblock \bibinfo{journal}{\emph{Proceedings of the ACM on Human-Computer
  Interactio, volume = {6}, number = {{CSCW2}}, pages = {1--24}, year = {2022},
  doi = {10.1145/3555762},}} (\bibinfo{year}{[n.\,d.]}).
\newblock


\bibitem[Sweeney(2002)]%
        {k-ano}
\bibfield{author}{\bibinfo{person}{Latanya Sweeney}.}
  \bibinfo{year}{2002}\natexlab{}.
\newblock \showarticletitle{k-Anonymity: A Model for Protecting Privacy}.
\newblock \bibinfo{journal}{\emph{International Journal of Uncertainty,
  Fuzziness and Knowledge-Based Systems}}  \bibinfo{volume}{10}
  (\bibinfo{year}{2002}), \bibinfo{pages}{557--570}.
\newblock


\bibitem[Valdez and Ziefle(2019)]%
        {ValdezZ19}
\bibfield{author}{\bibinfo{person}{Andr{\'{e}}~Calero Valdez} {and}
  \bibinfo{person}{Martina Ziefle}.} \bibinfo{year}{2019}\natexlab{}.
\newblock \showarticletitle{The users' perspective on the privacy-utility
  trade-offs in health recommender systems}.
\newblock \bibinfo{journal}{\emph{International Journal of Human-Computer
  Studies}}  \bibinfo{volume}{121} (\bibinfo{year}{2019}),
  \bibinfo{pages}{108--121}.
\newblock
\urldef\tempurl%
\url{https://doi.org/10.1016/j.ijhcs.2018.04.003}
\showDOI{\tempurl}


\bibitem[Wang et~al\mbox{.}(2020)]%
        {WangZFY20}
\bibfield{author}{\bibinfo{person}{Teng Wang}, \bibinfo{person}{Xuefeng Zhang},
  \bibinfo{person}{Jingyu Feng}, {and} \bibinfo{person}{Xinyu Yang}.}
  \bibinfo{year}{2020}\natexlab{}.
\newblock \showarticletitle{A Comprehensive Survey on Local Differential
  Privacy toward Data Statistics and Analysis}.
\newblock \bibinfo{journal}{\emph{Sensors}} \bibinfo{volume}{20},
  \bibinfo{number}{24} (\bibinfo{year}{2020}), \bibinfo{pages}{7030}.
\newblock
\urldef\tempurl%
\url{https://doi.org/10.3390/s20247030}
\showDOI{\tempurl}


\bibitem[Warner(1965)]%
        {RRT}
\bibfield{author}{\bibinfo{person}{Stanley~L. Warner}.}
  \bibinfo{year}{1965}\natexlab{}.
\newblock \showarticletitle{Randomized response: A survey technique for
  eliminating evasive answer bias}.
\newblock \bibinfo{journal}{\emph{J. Amer. Statist. Assoc.}}
  \bibinfo{volume}{60.309} (\bibinfo{year}{1965}), \bibinfo{pages}{63--69}.
\newblock


\bibitem[Xiong et~al\mbox{.}(2020)]%
        {xiong}
\bibfield{author}{\bibinfo{person}{Aiping Xiong}, \bibinfo{person}{Tianhao
  Wang}, \bibinfo{person}{Ninghui Li}, {and} \bibinfo{person}{Somesh Jha}.}
  \bibinfo{year}{2020}\natexlab{}.
\newblock \showarticletitle{Towards Effective Differential Privacy
  Communication for Users' Data Sharing Decision and Comprehension}. In
  \bibinfo{booktitle}{\emph{SP~'20: {IEEE} Symposium on Security and Privacy}}.
  \bibinfo{publisher}{IEEE}, \bibinfo{pages}{392--410}.
\newblock
\urldef\tempurl%
\url{https://doi.org/10.1109/SP40000.2020.00088}
\showDOI{\tempurl}


\end{thebibliography}

\section*{Appendix}
\appendix
\section{SURVEY DETAILS}
\label{appendix:survey}
In this appendix, we provide the description of our scenario, the explanations and questions on subjective and objective understanding.
The elements of the questions that are marked magenta indicate the changes that were implemented after our pilot study.




\subsection{Descriptions and Explanations}
\label{appendix:descriptions}
\subsubsection*{Scenario}
\label{appendix:scenario}
Consider a school survey on drug abuse including every student at the school. The answers of the students are stored in a database as shown in the table. Each row consists of the name, the age, and the class of the student as well as a value (“yes” or “no”) indicating whether the student uses drugs or not. In order to raise awareness, parents can query the database. At the same time, the school wants to protect the privacy of each student. To that end, the database is protected with a privacy mechanism. The students' privacy protection is explained to the parents and students.
		
Bob’s mother, Eve, finds out that there is high drug use in her son’s class.
She wants to know who---and especially if her son Bob---is using drugs.
From Bob’s class list, which includes the birth date of each classmate,
Eve knows the exact age of all students in Bob’s class.
Bob is the only student in his class who is $16.3$ years old.

\subsubsection*{$K$-Anonymity} 
$K$-anonymity provides privacy protection by generalizing or removing
		all sensitive columns of the database that might be used to re-identify a student.
		This way, for every student in the database, there is a group of at least $k$ students
		with the same answers in all sensitive columns.
		In other words, there are always at least $k$ indistinguishable students.
		Accordingly, $k$ is the privacy parameter, which determines the level of privacy protection.
		The higher the privacy parameter $k$, the more indistinguishable students exist in each group,
		resulting in a stronger privacy protection.
		
		Assume the school sets the privacy parameter to $k=4$,
		\ie, the database is modified in a way that it always yields
		at least $4$ indistinguishable students.
		Specifically, the database now contains four students from Bob’s class
		with a generalized age of $16$ (Peter, Bob, Marie, and Lucas).
		Names are not shown.
		Please note that Eve cannot link the individual rows to the respective students.
		Bob’s drug use could be indicated by each of the four rows equally likely.
		Hence, Bob---and its drug use---remain hidden in the group of four indistinguishable students.
		
\subsubsection*{Differential Privacy \RISK}
Differential privacy provides privacy protection
		by randomly modifying statistical results extracted from the database.
		The results therefore indicate the true answers of the students with a certain probability only:
		Every student in the database has a certain privacy risk
		that their true answer can be identified.
		This risk is controlled by a privacy parameter,
		which determines the level of privacy protection.
		A privacy parameter closer to zero reduces the privacy risk,
		resulting in a stronger privacy protection.
		
		Assume the school sets the privacy parameter in a way
		that yields a privacy risk for the students of $75$\,\%,
		\ie, the true answers of the students are indicated with a probability of $75$\,\%.
		Now, Eve accesses the database and asks for the number of $16.3$ years old drug-using students in Bob’s class.
		Please note that Eve does not know
		whether the returned result indicates Bob’s true answer or not.
		The result was modified and might therefore be false.
		Bob has a privacy risk of $75$\,\%.
		
\subsubsection*{Differential Privacy \RRT} 
Local differential privacy provides privacy protection
		by randomly modifying the students' answers containing sensitive information,
		before storing them in the database.
		The stored answers therefore correspond to the true answers of the students with a certain probability only:
		Every student in the database has a certain probability
		that their true answer is stored.
		This probability is controlled by a privacy parameter,
		which determines the level of privacy protection.
		A privacy parameter closer to zero reduces the probability,
		resulting in a stronger privacy protection.
		
		Assume the school sets the privacy parameter in a way
		that the mechanism stores the true answer of a student with a probability of $75$\,\%.
		Now, Eve accesses the database and asks for the number of 16.3 years old drug-using students in Bob’s class.
		Please note that Eve does not know
		whether the returned result indicates Bob's true answer or not.
		Bob's answer was modified and might therefore be false.
		Bob's true answer is stored with a probability of $75$\,\%. 

\subsubsection*{Differential Privacy \DEF} Differential privacy provides privacy protection
		by randomly modifying statistical results extracted from the database.
		The true results extracted from the students' answers are therefore returned with a certain probability only:
		The probability of the same result if one of the students gave a different answer has a certain difference to the probability of the true result.
		This difference is controlled by a privacy parameter,
		which determines the level of privacy protection.
		A privacy parameter closer to zero reduces the difference,
		resulting in a stronger privacy protection.
		
		Assume the school sets the privacy parameter in a way that
		the probability of returning any statistical result is $3$ times
		the probability of the same result if one of the students gave a different answer.
		Now, Eve accesses the database and asks for the number of $16.3$ years old drug-using students in Bob’s class.
		Please note that Eve does not know
		whether the returned result indicates Bob's true answer or not.
		The result was modified and might therefore be returned if Bob gave a different answer.
		The probability of the true result being returned
		is at most $3$ times the probability of the same result if Bob did not use drugs.

\begin{table}[tb]
	\caption*{Database with original answers}
	\label{tab:example1}
	\begin{tabularx}{\columnwidth}{XXXXX}
		\toprule
		row & name & class & age & drug use \\ 
		\midrule
		1 & Anna & c & 15.2 & false \\
		2 & Steve & c & 15.8 & true \\
		$\vdots$ & $\vdots$ & $\vdots$ & $\vdots$ & $\vdots$ \\
		9 & Peter & c & 16.0 & false \\
		10 & Bob & c & 16.3 & true \\
		11 & Marie & c & 16.6 & true \\
		12 & Lucas & c & 16.2 & true \\
		\bottomrule
	\end{tabularx}
\end{table}

\begin{table}[tb]
	\caption*{Anonymized database with $k=4$}
	\label{tab:example1-k}
	\begin{tabularx}{\columnwidth}{XXXX}
		\toprule
		row & class & age & drug use \\ 
		\midrule
		1 & c & 15 & false \\
		2 & c & 15 & true \\
		$\vdots$ & $\vdots$ & $\vdots$ & $\vdots$ \\
		9 & c & 16 & false \\
		10 & c & 16 & true \\
		11 & c & 16 & true \\
		12 & c & 16 & true \\
		\bottomrule
	\end{tabularx}
\end{table}

\subsection{Subjective Understanding Questions}

\textit{Answer the following questions on a scale from 1 (Not at all) to 7 (Very)}
\begin{itemize}
	\item \textit{How difficult is it to understand the \change{level of privacy protection provided by this} mechanism? (7=Very difficult, 1=Not at all difficult)}
	\item \textit{How do you rate your ability to explain the \change{level of privacy protection provided by} this mechanism to another person? (7=Very well, 1=Not at all well)}
	\item \textit{How well do you understand the \change{level of} privacy protection provided by this mechanism? (7=Very well, 1=Not at all well)}
	\item \textit{Your comments:}
\end{itemize}

\subsection{Objective Understanding Questions}
%
\begin{enumerate}
	\item \textit{What is more likely: that Eve assumes Bob uses drugs or that Eve assumes Bob does not use drugs?}
	\begin{itemize}
		\item \textit{That Eve assumes Bob uses drugs}
		\item \textit{That Eve assumes Bob does not use drugs}
		\item \textit{Both cases are equally likely}
	\end{itemize}
	\item \textit{Assume that the \change{privacy parameter} is higher. That implies …}
	\begin{itemize}
		\item \textit{The same \change{level of} privacy protection}
		\item \textit{A \change{stronger} privacy protection}
		\item \textit{A \change{weaker} privacy protection}
	\end{itemize}
	\change{
	\item \textit{Assume \change{that the privacy parameter is $0$.} That implies...}
	\begin{itemize}
		\item \textit{\change{That the mechanism provides all exact true data of all the students}}
		\item \textit{\change{That the mechanism provides only false data that is completely independent of the data of the students}}
		\item \textit{\change{That the mechanism does not provide any data at all}}
	\end{itemize}
	}
	\item \textit{Assume, Peter’s father wants to find out about Peter’s drug use. Peter has …}
	\begin{itemize}
		\item \textit{The same level of privacy protection \change{as Bob}}
		\item \textit{A stronger privacy protection \change{than Bob}}
		\item \textit{A weaker privacy protection \change{than Bob}}
	\end{itemize}
\end{enumerate}

\end{document}